\documentclass[]{spie}  %>>> use for US letter paper

\usepackage{amsmath,amsfonts,amssymb} \usepackage{graphicx}
\usepackage[colorlinks=true, allcolors=blue]{hyperref}
\usepackage{lineno}
\usepackage[loose]{units}
\usepackage[export]{adjustbox}
\usepackage{svg}

\newcommand\mumux{$\mu$mux}
\newcommand\smurf{SMuRF}

\newcommand{\Vth}{\hat{V}_\mathrm{th}}
\newcommand{\Isc}{\hat{I}_\mathrm{SC}}
\newcommand{\Iob}{\hat{I}_\mathrm{OB}}
\newcommand{\Zeq}{\hat{Z}_\mathrm{eq}}
\newcommand{\Ztes}{\hat{Z}_\mathrm{TES}}
\newcommand{\taueff}{\tau_\mathrm{eff}}
\newcommand{\LI}{\mathcal{L}_I}
\newcommand{\Rshunt}{R_\mathrm{sh}}

\newcommand{\Ites}{\hat{I}_\mathrm{TES}}

\newcommand{\orcid}[1]{\href{https://orcid.org/#1}{\includegraphics[height=2ex]{ORCIDiD_icon64x64.png}}}

\title{
The Simons Observatory: Complex Impedance Measurements for a Full Focal-Plane Module
}

\author[1]{Jack Lashner}
\author[2]{Joseph Seibert}
\author[2]{Max Silva-Feaver}
\author[3]{Tanay Bhandarkar}
\author[4]{Kevin T. Crowley}
\author[5]{Shannon M. Duff}
\author[6]{Daniel Dutcher}
\author[7]{Kathleen Harrington}
\author[8]{Shawn W. Henderson}
\author[1]{Amber D. Miller}
\author[9, 10]{Michael Niemack}
\author[6]{Suzanne Staggs}
\author[6]{Yuhan Wang}
\author[6]{Kaiwen Zheng}

\affil[1]{Department of Physics and Astronomy, University of Southern California, Los Angeles, CA 90007, USA}
\affil[2]{Department of Physics, University of California, San Diego, La Jolla, CA 92093, USA}
\affil[3]{Department of Physics and Astronomy, University of Pennsylvania, 209 S 33rd St. Philadelphia, PA 19104, USA}
\affil[4]{Department of Physics, University of California, Berkeley, CA 94720, USA}
\affil[5]{National Institute of Standards and Technology, Boulder, Colorado 80305, USA} % Princeton
\affil[6]{Joseph Henry Laboratories of Physics, Jadwin Hall, Princeton University, Princeton, NJ 08544, USA}
\affil[7]{Department of Astronomy and Astrophysics, University of Chicago, Chicago, IL, USA}
\affil[8]{SLAC National Accelerator Laboratory, Menlo Park, CA 94025, USA}
\affil[9]{Department of Physics, Cornell University, Ithaca, NY 14853, USA}
\affil[10]{Department of Astronomy, Cornell University, Ithaca, NY 14853, USA}

\authorinfo{Further author information: (Send correspondence to Jack Lashner)\\
Jack Lashner: E-mail: lashner@usc.edu}

\begin{document} 
\maketitle

\begin{abstract}

The Simons Observatory (SO) is a ground based Cosmic Microwave Background
experiment that will be deployed to the Atacama Desert in Chile.
SO will field over 60,000 transition edge sensor (TES) bolometers that will
observe in six spectral bands between 27 GHz and 280 GHz with the goal of
revealing new information about the origin and evolution of the universe. 
SO detectors are grouped based on their observing frequency and packaged into
Universal Focal Plane Modules, each containing up to 1720 detectors
which are read out using microwave SQUID multiplexing and the SLAC
Microresonator Radio Frequency Electronics (\smurf).
By measuring the complex impedance of a TES we are able to access many
thermoelectric properties of the detector that are difficult to determine using
other calibration methods, however it has been difficult historically to
measure complex impedance for many detectors at once due to high sample rate
requirements.
Here we present a method which uses \smurf\ to measure the complex impedance of
hundreds of detectors simultaneously on hour-long timescales.
We compare the measured effective thermal time constants to those estimated
independently with bias steps.
This new method opens up the possibility for using this characterization tool
both in labs and at the site to better understand the full population of SO
detectors.

% The Simons Observatory (SO) is a ground based Cosmic Microwave Background
% experiment that is scheduled to deploy to the Atacama Desert in Chile in 2022.
% SO will field over 60,000 transition-edge sensor (TES) bolometers that will
% observe in six spectral bands between 27 GHz and 280 GHz.
% Here we present a method which uses \smurf\ to measure the complex impedance of
% hundreds of detectors simultaneously on hour-long timescales.
% We compare the measured effective thermal time constants to those estimated
% independently with bias steps.
% This new method opens up the possibility for using this characterization tool
% both in labs and at the site to better understand the full population of SO
% detectors.

\end{abstract}

\section{Introduction}

The Simons Observatory (SO) is a new Cosmic Microwave Background (CMB)
experiment being built on Cerro Toco in Chile, which will make precision
measurements of the temperature and polarization anisotropies of the CMB in
order to measure and constrain fundamental properties of the Universe
\cite{adeSimonsObservatoryScience2019}.
SO will field over 60,000 Transition Edge Sensor (TES) detectors, split between
a 6 m crossed Dragone large aperture telescope
(LAT)\cite{zhuSimonsObservatoryLarge2021, parshleyOpticalDesignSixmeter2018}
containing 30,000+ detectors, and an array of three 42~cm small aperture
refracting telescopes (SATs) containing a total of 30,000 detectors
\cite{galitzkiSimonsObservatoryInstrument2018, zhuSimonsObservatoryLarge2021,
aliSmallApertureTelescopes2020}.
SO will use dichroic pixels observing in six spectral bands, at frequencies
27/39~GHz (LF), 93/145~GHz (MF) and 225/280~GHz (UHF)
in order to measure the CMB and remove foreground contamination.
% Changed freqs from:
% \unit[30/40]{GHz} (LF), \unit[90/150]{GHz} (MF) and \unit[230/280]{GHz} (UHF)

Advancements in readout architecture are essential to accommodate the large
detector counts of SO and future CMB experiments.
High multiplexing factors are necessary to decrease the thermal loading on the
focal-plane, and decrease the complexity of the readout wiring.
SO will use microwave superconducting quantum interference device (SQUID)
multiplexing (\mumux) \cite{matesDemonstrationMultiplexerDissipationless2008},
where TESs are inductively coupled to superconducting microresonators between
4-6~GHz via RF SQUIDs.
The SQUID coupling transduces current through the TES into a change in
effective inductance of the microresonator, causing a measurable shift in its
resonance frequency.
At room temperature we use SLAC Microwave RF (\smurf) electronics
\cite{hendersonHighlymultiplexedMicrowaveSQUID2018} to generate the tones used
to interrogate resonators and track their resonance frequencies.
In addition, \smurf\ provides DC TES bias voltages, and the flux-ramp signal
which is used to linearize the RF SQUID response.

The optical coupling, detector arrays, and cold electronics are all packaged
together into 150~mm hexagonal assemblies called universal focal-plane modules
(UFM) \cite{mccarrickSimonsObservatoryMicrowave2021}.
The focal planes of each telescope will be composed of tiled UFMs, with each
SAT focal-plane containing 7 modules, and each of seven LAT optics tubes
containing 3 modules.
UFMs can support up to 1820 multiplexing channels and up to 1728 optical
detectors, and will be read out using two Radio Frequency (RF) transmission
lines.
The UFM also provides routing for the 12 DC bias lines used to bias to the
TESs, and two flux ramp lines used to linearize the SQUID response
\cite{matesFluxRampModulationSQUID2012}.

Before deployment, each UFM is tested and characterized in an SO testbed
\cite{wangSimonsObservatoryFocalPlane2021}.
By measuring detector noise, current-voltage curves (IVs), and measuring the
change in current through the TES in response to a small step in DC bias
voltage (bias steps), we are able to determine detector properties such as the
normal TES resistance ($R_N$), the bias power required to drive detectors to
90\% of $R_N$ ($P_\mathrm{SAT}$), the thermal conductance $G$, and the
effective thermal time constant $\taueff$.
Some detector parameters such as the logarithmic sensitivities to temperature
and current ($\alpha$ and $\beta$ respectively), the loop-gain $\LI$, and the
TES heat capacity $C$ are difficult to measure independently from each other due to degeneracy.
The complex impedance (CI) of the TES $\Ztes$ provides a method by which we can
access these parameters by measuring the response of the TES to sinusoidal bias
voltage stimuli. 
CI measurements have previously been performed on SO prototype bolometers
\cite{cothardComparingComplexImpedance2020}, however these measurements have
been limited to only a few bolometers due to the high sample rate requirements.
The high bandwidth of \smurf\ allows us to take data at sample rates of up to
25~kHz for over 150~detectors at a time, and the Digital to Analog Converters
(DACs) that power the bias lines can be used to send sine waves along the TES
bias lines at frequencies approaching 5~kHz.
\smurf\ allows us to measure the complex
impedance of large batches of TESs simultaneously, and perform preliminary and
in-situ array-scale characterization in a matter of hours.

In this paper, we describe a procedure of measuring the complex impedance of SO
detectors at array scales using the \smurf\  readout electronics.
Section 2 presents the TES model used to interpret the CI measurements.
Section 3 describes the methods used to characterize detectors with complex
impedance measurements, and bias step measurements.
Section 4 presents the complex impedance results for a prototype UFM,
and compares the effective thermal time constants with those calculated based on
bias steps.

%%%%%%%%%%%%%%%%%%%%%%%%%%%%%%%%%%%%%%%%%%%%%%%%%%%%%%%%%%%%%%%%%%%%%%%%%%%%%%%%
% Section 2: Background
%%%%%%%%%%%%%%%%%%%%%%%%%%%%%%%%%%%%%%%%%%%%%%%%%%%%%%%%%%%%%%%%%%%%%%%%%%%%%%%%
\section{Background}
\label{sec:background}

% TES Schematic fig {{{1
\begin{figure}[h]
  \begin{center}
    \includegraphics[width=0.5\textwidth]{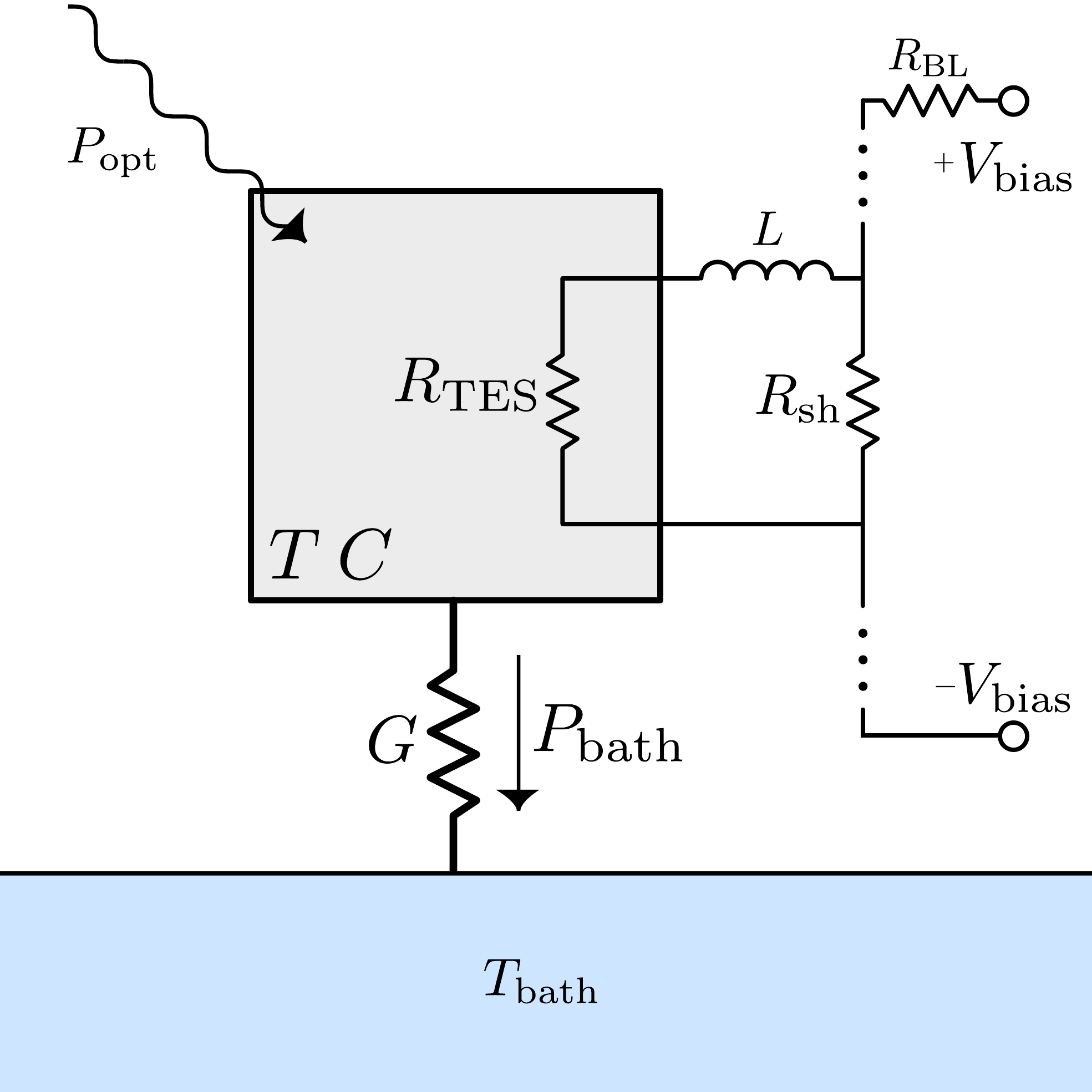}
  \end{center}
  \caption{
  Diagram of the bolometer model used in this study.
  The TES is modeled as an island thermally isolated from the bath with heat
  capacity C coupled to a bath of temperature $T_\mathrm{bath}$ with thermal
  conductance $G$.
  Each bias line contains hundreds of TES circuits in series, each with the
  TES wired in parallel with a shunt resistor.
  }
  \label{fig:tes_schematic}
\end{figure}
% }}}1

To model the complex impedance of the TES, we follow the formalism used by
Irwin and Hilton \cite{irwinTransitionEdgeSensors2005} and model the TES
bolometer as a single thermal block with temperature $T$ and heat capacity $C$.
This is a very simplified model, but measuring the complex impedance of the TES
can help us understand departures from these assumptions.
The TES is coupled to a thermal bath of temperature $T_b$ with thermal
conductance $G = \left.\partial P_\mathrm{bath} / \partial T \right|_{T_b}$.
We bias the detectors using a constant-voltage bias $V_\mathrm{bias}$ that is
applied to each bias line, which consists of a large bias line resistor
$R_\mathrm{BL}\approx \unit[16]{k\Omega}$ in series with around 150 TES circuits.
Each TES circuit is modeled as the TES resistance in series with a parasitic
inductance $L$, and with a shunt resistor $\Rshunt \approx 0.4$ m$\Omega$ connected
in parallel.
A depiction of this model can be seen in Figure \ref{fig:tes_schematic}.

We substitute the full bias schematic shown in Figure \ref{fig:tes_schematic}
with a Thevenin equivalent circuit containing just a single TES, with
equivalent voltage $\Vth$ and equivalent impedance $\Zeq$
where both are complex phasors that depend on the excitation frequency $\omega$.
For the bias circuit described above, the equivalent impedance at low
frequencies is given by 
$Z_\mathrm{eq}(\omega) \approx \Rshunt + i \omega L$.
The dynamics of the TES are governed by the set of coupled differential
equations:

% Electrothermal eqs {{{1
\begin{equation}
  C \frac{d T}{dt} = - P_\mathrm{bath} + P_J + P_\mathrm{opt}
  \label{eq:theramal_eq}
\end{equation}
\begin{equation}
  L \frac{d I}{dt} = V_\mathrm{th} - I \Rshunt - I R(T, I)
  \label{eq:electro_eq}
\end{equation}
% }}}1

where $P_\mathrm{bath}$ is the power flowing from the TES to the thermal bath,
$P_J$ the power from joule heating, $P_\mathrm{opt}$ the power deposited
onto the TES through optical coupling, and $R(T, I)$ the TES resistance.

Under the constant-voltage bias, in-transition detectors are able to achieve
negative electrothermal feedback which keeps the TES in transition.
The detector is characterized by its logarithmic temperature sensitivity, logarithmic current sensitivity, and the loop-gain:
\begin{equation}
  \alpha \equiv \left.\frac{\partial \log R}{\partial \log T}\right|_{I_0}
    \qquad
  \beta \equiv \left.\frac{\partial \log R}{\partial \log I}\right|_{T_0}
    \qquad
  \LI \equiv \frac{P_{J_0} \alpha}{G T_0}
  \label{eq:sensitivies}
\end{equation}
where $I_0$, $T_0$, and $P_{J_0}$ are the equilibrium current, bolometer temperature
and bias power.

For small perturbations $\delta I$ and $\delta T$, equations \ref{eq:theramal_eq} and \ref{eq:electro_eq}
can be linearized and solved\cite{beckerSUBMILLIMETERVIDEOIMAGING}.
These solutions contain two time constants: $\tau_\mathrm{el}$ characterizing
the initial electrical response, and $\taueff$ characterizing the effective
thermal decay to equilibrium.
Both time constants can be seen in a TES's response to a sudden voltage bias
step, as is shown in Figure \ref{fig:bias_steps}.
SO bolometers have a very small inline inductance $L$, which causes $\taueff$
to be much longer than $\tau_\mathrm{el}$ and makes $\taueff$ the determining
factor for how fast the TES can respond to an optical signal.

Solving the coupled differential equation yields
\begin{equation}
  \taueff = \tau_0
  \left(1 + 
    \frac{(1 - \Rshunt/R) \LI}{1 + \beta + \Rshunt/R}
  \right)^{-1}
  \label{eq:tau_eff_def}
\end{equation}
where $\tau_0 = C / G = \tau_I (1 - \LI)$ is the natural thermal time constant in
the absence of electrothermal feedback and $\tau_I$ is the thermal time
constant in the case of a constant-current bias source.

The complex impedance of the TES can be extracted from the general solutions,
and is given by
\begin{equation}
  \Ztes(\omega) = R(1 + \beta) + 
  \frac{R \LI}{1 - \LI}
  \frac{2 + \beta}{1 + i \omega \tau_I}.
  \label{eq:ztes_det_params}
\end{equation}
where $\omega$ is the excitation frequency.
$\Ztes$ can be accessed by measuring the transfer function across the TES
as is described in the next section.

\section{Methods}
\label{sec:methods}

The measurements in this section were carried out on a prototype SO UFM
in a testbed designed for optical characterization of SO detectors
\cite{seibertDevelopmentOpticalDetector2020}.
Two thirds of the detectors in the UFM are covered by a gold-plated silicon
mask allowing us to test both optical and dark detectors simultaneously.
IVs taken immediately prior are used to determine parameters such as $R_N$.
The DC voltages for each TES bias line were chosen based on IVs to maximize the
number of detectors with resistances in the range 30\% - 60\% $R_N$.
Note that the measurements shown in this section demonstrate a useful
characterization method, but do not reflect expected operating parameters.
The measurements were taken at a higher bath temperature (\unit[150]{mK}) than
that expected during site observation (\unit[100]{mK}), and this particular set
of TESs have optical efficiency that is too low to be deployed for observations.

\subsection{Complex Impedance Measurement}
\begin{figure}[t]
  \begin{center}
    \includegraphics[
    width=0.95\textwidth,
    trim=120 0 120 0
    ]{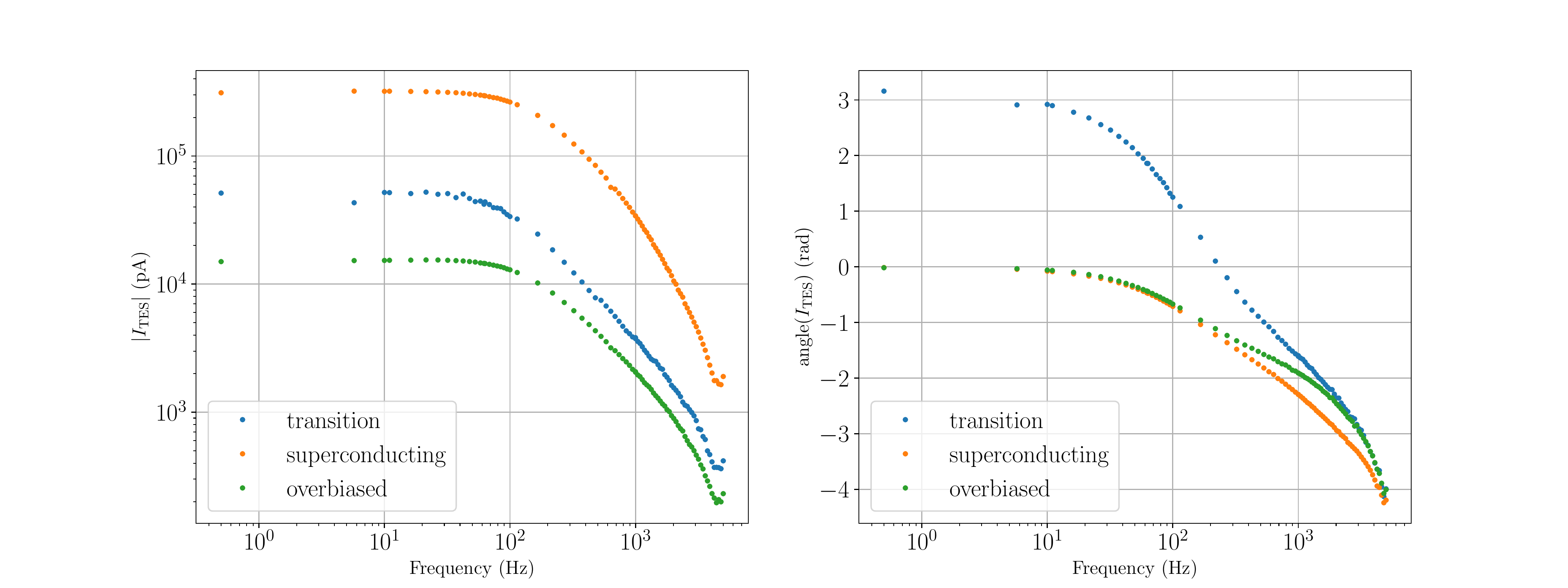}
  \end{center}
  \caption{
    Magnitude (left) and phase (right) of $\Ites(f)$ for a single detector, where $f$ is the excitation frequency.
    The magnitude of $\Ites$ is the amplitude of the sine wave through the
    detector, and the phase is measured relative to the commanded bias waveform. 
    We show the TES current measured while the detector is superconducting, 
    overbiased, and in transition.
    The small uptick at high frequencies, is likely due to limitations in the
    \smurf\ waveform generation, but this is absorbed into $\Vth$ in
    \ref{eq:vth_meas} and is not seen in the measured $\Ztes$.
  }
  \label{fig:CI_Ites}
\end{figure}
\begin{figure}[ht]
  \begin{center}
    \includegraphics[
      width=0.45\textwidth,
      trim=90 0 20 0
    ]{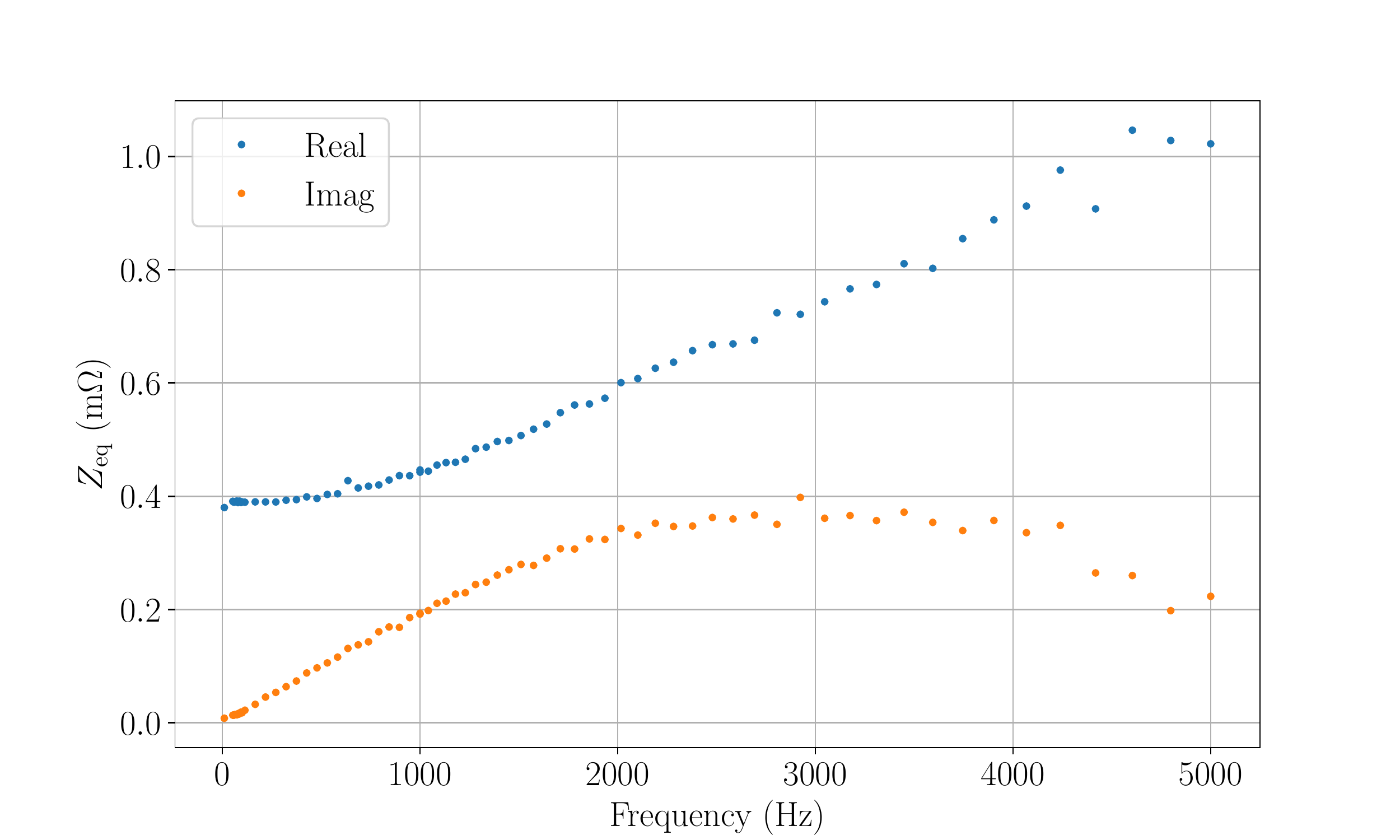}
    \includegraphics[
      width=0.45\textwidth,
      trim=20 0 90 0
    ]{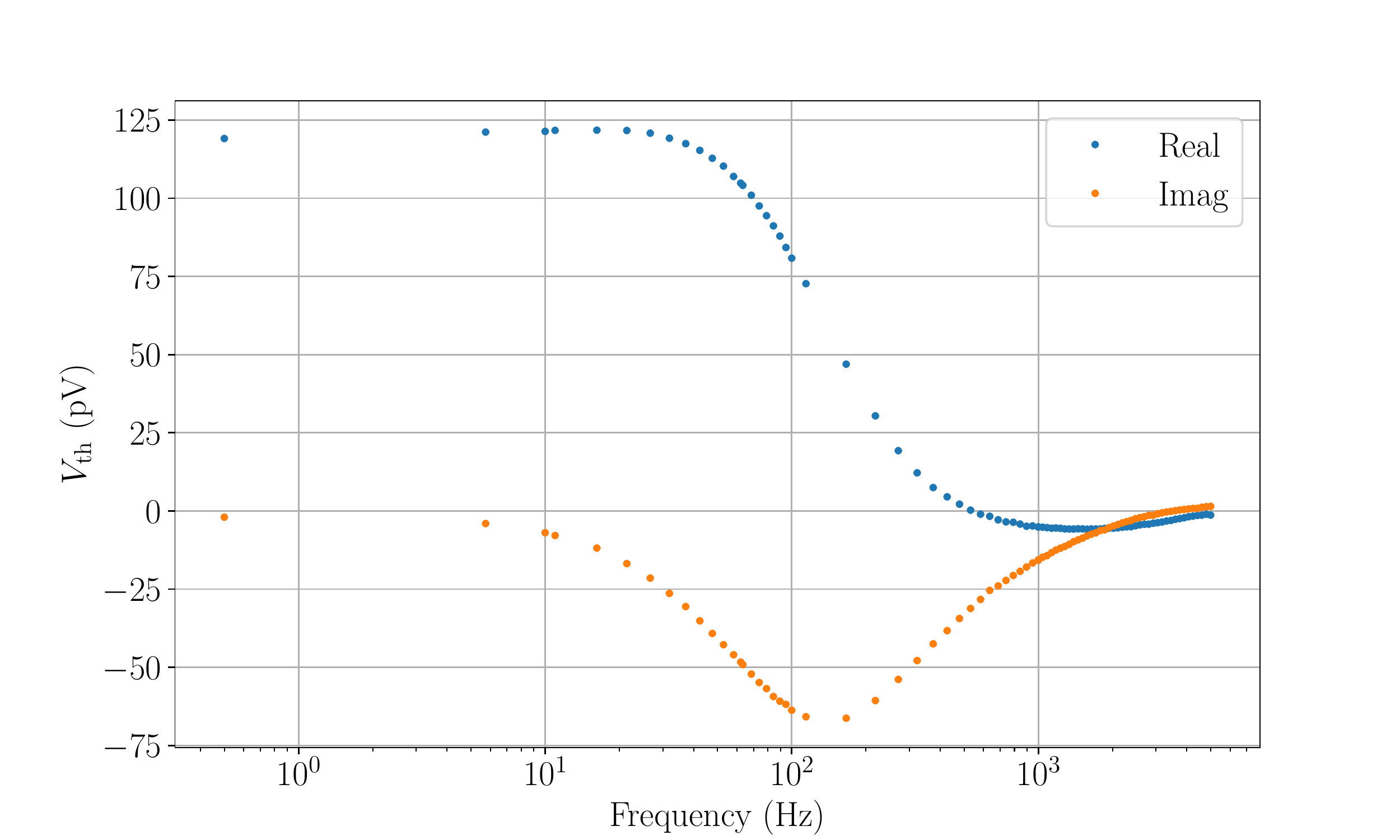}
  \end{center}
  \caption{
    $\Zeq(f)$ (left) and $\Vth(f)$ (right) for a single detector, determined by
    the TES response to sinusoidal stimuli in the overbiased and superconducting
    states.
    In each plot the real part is shown in blue, and the imaginary part is
    shown in orange.
    Using the simple electrical model of the bias line described in Section
    \ref{sec:background}, at low excitation frequencies we expect the equivalent
    impedance to approach $\Zeq(\omega) \approx \Rshunt + i \omega L$.
    In the left plot we are able to observe $R_\mathrm{sh} \approx 0.4$
    m$\Omega$ from the low-frequency limit of $\mathrm{Re}[\Zeq]$,
    and estimate the inductance $L$ as the slope of the low-frequency
    limit of $\mathrm{Im}[\Zeq]$.
    $\Zeq(f)$ is plotted against a linear axis so we are able to more clearly show
    the linear dependence of $\mathrm{Im}[\Zeq]$ on frequency.
    % \vspace*{5in}
  }
  \label{fig:zeq_and_vth}
\end{figure}

The transfer function of the TES in response to an excitation frequency
$\omega$ is measured by sending a small sine wave along the bias line on top of
the DC bias voltage biasing the TESs into transition.
For each data sample, the commanded output of the DACs driving the TES bias line
is recorded along with the TES response.
We use these commanded TES bias values as a reference for a digital lock-in procedure\cite{probstLowFrequencyDigital1985} that allows us to 
extract the amplitude and phase of the TES response relative to the commanded
bias with high levels of signal-to-noise.
We measure the transfer function while the detectors are superconducting,
overbiased, and in transition which can be seen in Figure \ref{fig:CI_Ites}.
The superconducting and overbiased measurements allow us to estimate the
Thevenin equivalent voltage and impedance of the system for every TES bias
circuit at each frequency \cite{lindemanComplexImpedanceMeasurements2007}
and thus correct for stray impedances in the system.
$\Vth$ and $\Zeq$ are given by
\begin{equation}
  \Vth(\omega) = \frac{R_N}{\Iob^{-1}(\omega) - \Isc^{-1}(\omega)}
  \label{eq:vth_meas}
\end{equation}
\begin{equation}
  \Zeq(\omega) = \Vth(\omega) / \Isc(\omega).
  \label{eq:zeq_meas}
\end{equation}
and the measured values for a single detector can be seen in Figure
\ref{fig:zeq_and_vth}.

\begin{figure}[htbp] \begin{center}
    \includegraphics[
      width=0.8\textwidth,
      trim=220 0 200 0
    ]{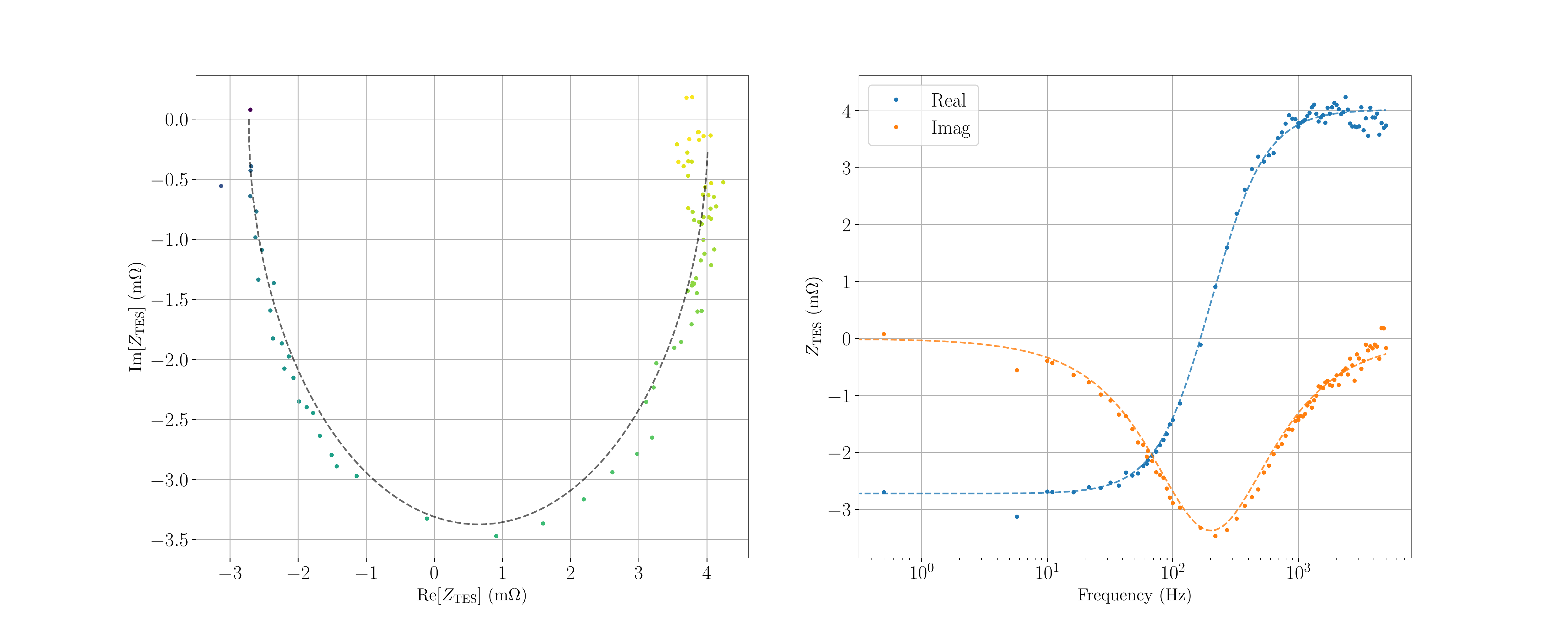}
  \end{center}
  \caption{
    $\Ztes$ of a single bolometer on the transition.
    On the left we see the semicircle that $\Ztes$ traces in the complex plane,
    where the color of the data points corresponds to the excitation frequency.
    On the right we see the real and imaginary components of $\Ztes$ both
    plotted against frequency.
    In both plots, the dashed lines show the $\Ztes$ fitted with equation
    \ref{eq:ztes_det_params}.
  }
  \label{fig:ztes}
\end{figure}

The complex impedance of the TES is then given by
\begin{equation}
  \Ztes(\omega) = \Vth(\omega) / \hat{I}(\omega)  - \Zeq(\omega)
  \label{eq:ztes_def}
\end{equation}
where $\hat{I}$ is the complex phasor representing the TES current in
transition.
$\Ztes$ for a single detector can be seen in Figure \ref{fig:ztes}.
The measured $\Ztes$ for each detector is then fit using the model in equation
\ref{eq:ztes_det_params} for the parameters $R$, $\LI$, $\beta$ and $\tau_I$,
which are used to compute derived parameters such as $\taueff$ and $\tau_0$.

Bipolar sine waves between 1 Hz and 5 kHz are applied to each bias line on top
of the DC TES bias voltage, with an amplitude of around 0.1\% of the DC bias
level.
The frequencies of the sine waves are limited by how quickly the two DACs that
generate the TES bias voltages can be programmed.
With the default \smurf\ 1~MHz clock-rate, the bias voltage can be updated at around 16~kHz.
This update rate can be increased by either increasing the clock-rate, or by
only sending a sine wave on a single DAC while keeping the other fixed.
As we approach excitation frequencies of 5 kHz the DACs can only be updated a
few times per cycle, however using digital lock-in procedure we are still
able to extract amplitudes and phases from the TES response.
There is a small amount variation in the amplitude of the output sine wave at
higher frequencies as can be seen in Figure \ref{fig:CI_Ites},
however, as shown in Figure~\ref{fig:zeq_and_vth} this is absorbed into $\Vth$
and does not effect the $\Ztes$ measurement.
$\Ztes$ approaches negative $R$ at low frequencies and $R (1 + \beta)$ at high
frequencies, so high fidelity measurements out to high frequencies are
particularly important for constraining $\beta$.
% When streaming 150 detectors at a time we are able to reach sampling rates 
% of 25~kHz, well above the Nyquist frequency required to determine the amplitude 
We sample detectors at 25~kHz, well above the Nyquist rate of the 5~kHz
excitation.

% Max's paragraph
% In previous readout implementations, complex impedance measurements on an
% entire array at once were not possible due to 1) insufficient bandwidth in the
% TES bias DACs in order to generate sine wave frequencies to constrain $\beta$
% and 2) insufficient data stream bandwidths to support streaming an arrays worth
% of data to disk at sample rates high enough to sample the highest frequency
% sine waves.
% This meant that this measurement was typically performed on a single channel at
% a time and required an additional external function generator and bias circuit
% which restricted this to a diagnostic check on one or two detectors as opposed
% to a wafer scale measurement that could provide statistics and spatial
% distributions of TES parameters like $\beta$.
% The SMuRF however has a TES bias DAC and filtering circuit that can be switched
% to a lower filtering higher bandwidth mode and support sine waves up to ~1kHz
% and a PCIe card streaming interface that supports 10 GB/s data rate per 4 GHz
% of RF readout (2000 Channels) bandwidth up to 60 GB/s streamed to a single
% server.

Performing this measurement on a full UFM takes around 30 minutes, with the
measurement time being limited by the fact that we need to measure the TES
response in 12 batches, corresponding to the 12 bias lines on our UFM.
The measurement is bottlenecked by the low-frequency data points, which
require more time due to their longer periods.
The measurement time can be shortened by reducing the number of samples at low
excitation frequencies, and by reducing the number of periods we measure for
those particular samples.
It is sufficient to take superconducting and overbiased measurements once per
array, and those measurements can be used for all in-transition measurements.
This can be run in parallel across multiple UFMs simultaneously, resulting in an
estimated measurement time for 30,000 detectors of approximately 2 hours.

Due to performance issues in the cryostat used to take these measurements, it
was not possible to heat the array to perform tests at multiple bath
temperatures.
This causes certain parameters such as $\LI$ and $\tau_0$ to be poorly
constrained, and could cause degeneracy between parameters such as $\beta$ and
$\tau_I$.
One could improve these measurements by taking data at multiple bath
temperatures and bias levels, forcing $\tau_0$ to remain constant across
datasets.

\subsection{Bias Step Measurement}
\label{sec:bias_steps}

\begin{figure}[h!]
  \begin{center}
    \includegraphics[width=0.6\textwidth]{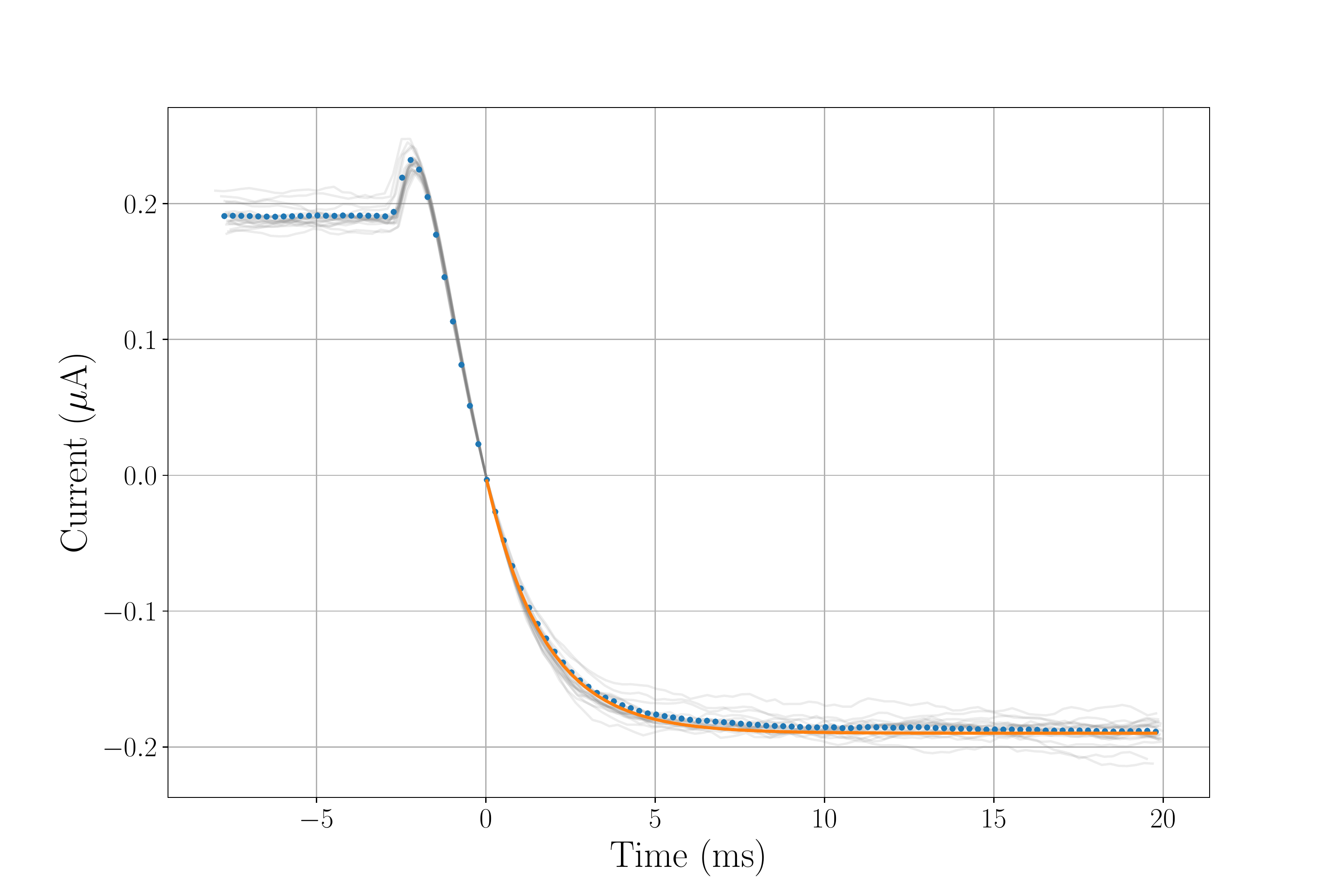}
  \end{center}
  \caption{
    TES response to bias steps taken for a single detector biased in the
    transition.
    This plot shows forty steps taken in quick succession, aligned such that
    the step midpoint occurs at $t=0$.
    Only the downward steps are shown here.
    The small increase in current before the step shows the initial electrical
    response of the TES, while the remainder of the step shows the slower
    thermal relaxation to equilibrium.
    Here the light grey lines show the individual steps, the blue points show 
    the average of all steps after they have been aligned, and the orange line
    shows the fit to the average step, starting at $t=0$ to avoid the initial
    electrical response.
  }
  \label{fig:bias_steps}
\end{figure}

We compare the $\taueff$ measured using complex impedance to those
independently measured from bias steps taken immediately before.
Bias step measurements consist of biasing the detectors into their transition,
and then sending a small-amplitude square wave on top of the DC bias voltage
\cite{koopmanAdvancedACTPolLowFrequency2018,
cothardComparingComplexImpedance2020}.
We perform many steps in quick succession, and align each step such that it crosses the midpoint at $t=0$.
We then use the aligned steps to compute the mean response for each TES.
We fit the thermal decay of the mean step response with a single-pole
exponential of the form $f(t)= A \exp(-t / \tau) + B$.
We only fit the response after $t=0$ to avoid the initial electrical response of the TES.
In Figure \ref{fig:bias_steps} we show each of the steps after alignment,
the mean step, and the exponential fit for a single TES.

To command the bias steps, we use \smurf's arbitrary waveform generator
to play a square wave with a period of 50 ms, and amplitude of around 1\% of
the DC bias voltage.
We take data on one bias line at a time to avoid crosstalk between bias lines,
and sample detectors at 4~kHz.
We apply around 40 individual steps to improve the statistical significance of
our measurements.
This measurement takes around 30 seconds for a full UFM, making it a very
useful tool for characterizing detectors throughout an observation.

\section{Results}

\begin{figure}[h!tpb]
  \begin{center}
    \includegraphics[width=0.95\textwidth]{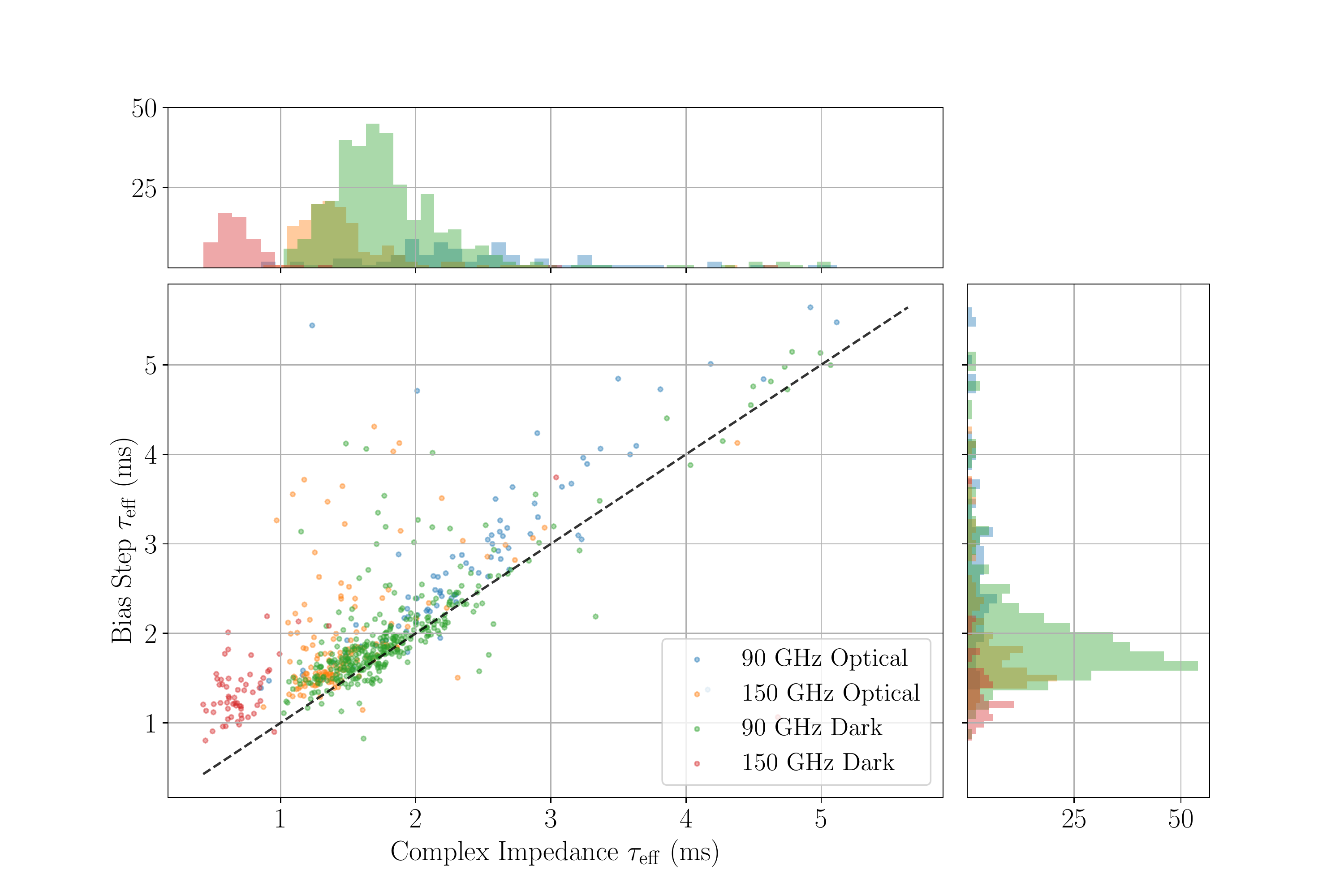}
  \end{center}
  \caption{
    Comparison of $\taueff$ from the complex impedance measurements with
    $\taueff$ estimated from bias steps.
    Each point represents a pair of measurements for a single detector.
    This plot shows dark and optical detectors in the 90~GHz and
    150~GHz bands.
    150~GHz detectors have a higher saturation power than the 90~GHz
    detectors, and we expect them to have faster time constants.
    Due to the elevated bath temperature, the majority of the optical 90~GHz
    detectors and many of the optical 150~GHz detectors are saturated and
    cannot be biased into the transition.
    The dashed black line is a guide line with a slope of one to show where
    the measurements would be equal.
  }
  \label{fig:bs_ci_comparison}
\end{figure}

We performed the measurements described in Section \ref{sec:methods} on an
early prototype UFM, and obtained data for 663 TESs.
This population includes both 90 and 150 GHz detectors, and includes both
detectors that are blocked off at the UFM, and those that have an optical
pathway out to room temperature.
This is a relatively low yield for a full UFM, partially due to the elevated
bath temperatures causing most of the optical 90~GHz detectors to be saturated,
but it is enough to observe population statistics and trends.
Figure \ref{fig:bs_ci_comparison} shows a comparison
between $\taueff^\mathrm{BS}$ and $\taueff^{CI}$ for each channel
that passed data quality cuts. 
In general we see good agreement between the two datasets,
though bias step calculations slightly overestimate $\taueff$ compared to the
complex impedance estimation, with this bias being worse for faster detectors.
It is possible that this bias stems from inadequacies of using a single-pole
exponential to fit the bias data, as it has been seen that the fitted $\taueff$
is somewhat sensitive to our choice of $t=0$ \cite{niemackDarkEnergyDesign2008}.
% This same bias can be seen in the characterization of prototype bolometers
% presented in Cothard 2020\cite{cothardComparingComplexImpedance2020}, and is in
% agreement with optical measurements performed on other SO testbeds.

\begin{figure}[h]
  \begin{center}
    \includegraphics[
        width=0.95\textwidth,
        trim=100 50 100 80]{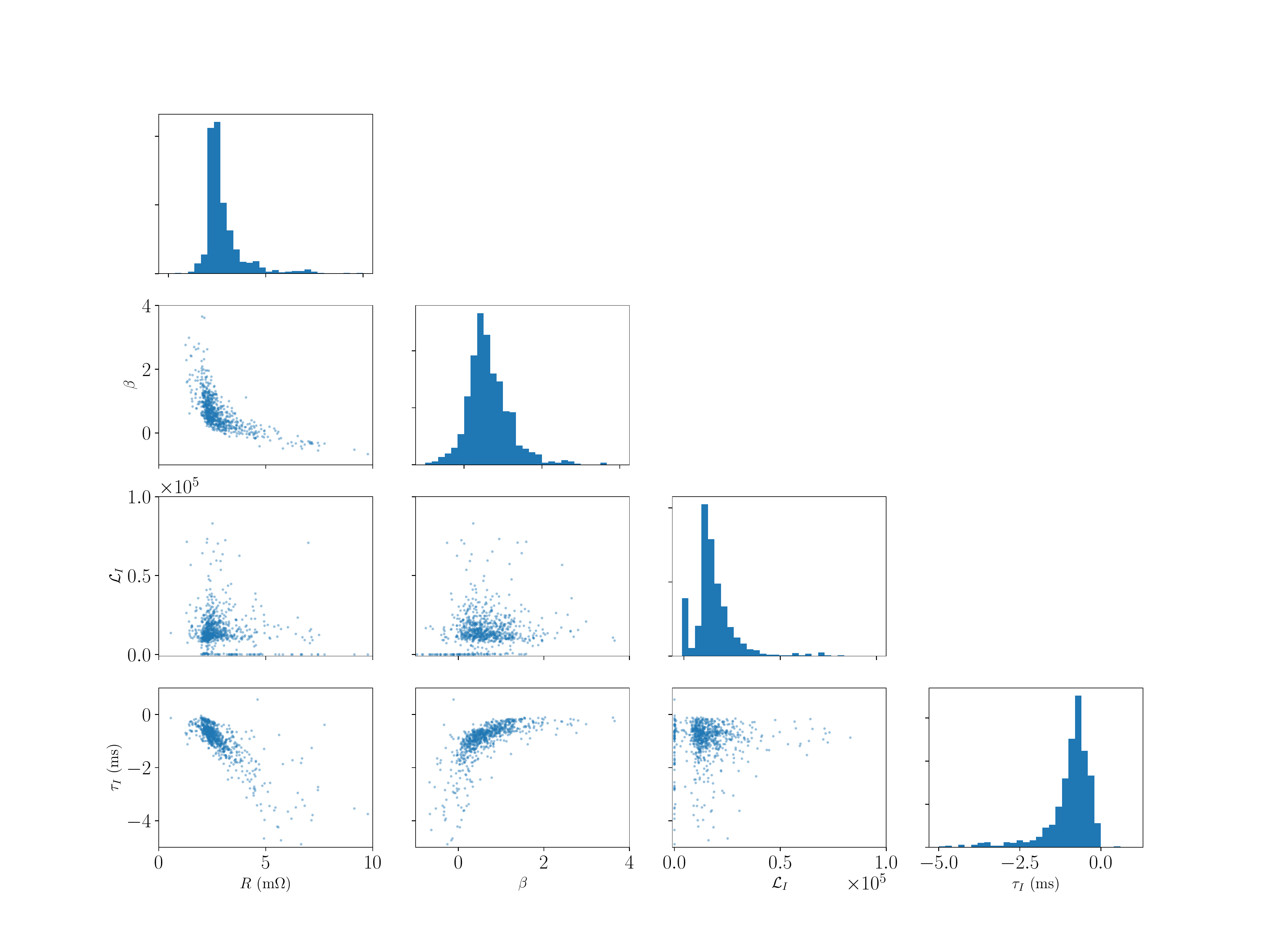}
  \end{center}
  \caption{
    Detector parameters fit to $\Ztes$ measurements, with each point
    representing the fit for a separate detector.
    Loop-gain fits are unphysically large, as they are not well constrained
    by measurements taken at a single bath temperature.
  }
  \label{fig:fit_degen}
\end{figure}

Figure \ref{fig:fit_degen} shows all CI fit parameters for each channel, and
how they compare to one another.
From these fits, it is clear that the loop-gain $\LI$, and thus the derived
parameter $\tau_0$, are poorly constrained by this dataset.
For optimally biased detectors, we expect large values of $\LI$, at least greater than 10. 
In this regime the loop-gain has a negligible effect on the complex impedance
in equation \ref{eq:ztes_det_params} compared to $\beta$ and $\tau_I$, and so
fits tend to gravitate towards the $\LI \rightarrow \infty$ limit.
% If we force $\LI$ to have a value around what we expect for SO detectors, for
% instance $\LI=20$, the value of $\taueff$ decreases by only a few percent
% compared to the infinite $\LI$ case.
% Since this causes $\taueff^{CI}$ to be smaller instead of larger, it does not
% account for the bias between complex impedance and bias step data.

A better estimation of $\LI$ could be achieved by repeating this
measurement at various bath temperatures, fixing $\tau_0$ across the datasets.
Alternatively, one could fix the thermal coupling coefficient $G$ to values
measured by taking IVs at various temperatures
\cite{wangSimonsObservatoryFocalPlane2021}.

\section{Conclusion}

In this work we present complex impedance results for 663 TESs on a prototype
UFM.
We compare the effective thermal time constant $\taueff$, determined from the
complex impedance measurement, with those estimated using bias-steps.
We observe that, though there is general agreement between the two methods,
bias-steps slightly overestimate $\taueff$ relative to complex impedance
measurements.
We also observe that certain parameters such as $\LI$ are not well constrained
by these complex impedance measurements.
This is unsurprising because we were operating at high bath temperatures due to
performance issues of our cryostat.

This work demonstrates a new implementation for characterizing TESs enabled by advances in readout technology
that allows us to take in-situ complex impedance measurements of thousands of
detectors using \smurf\ electronics.
This allows us to study population statistics for detector parameters that have
been historically difficult to measure, which can inform future detector
design.
Because this method does not require specialized hardware, it can be run at the
site to monitor parameters so they can be accessed by the SO analysis pipeline
for improved mapping.
In the future we plan on optimizing this measurement so it takes less time,
and retaking this data for SO deployment UFMs at operational temperatures to
better constrain these parameters, and to explore how SO bolometers deviate
from the simple thermal model presented here. 

\section{Acknowledgements}
This work was supported in part by a grant from the Simons Foundation (Award \#457687, B.K.).

% References
\bibliography{report} % bibliography data in report.bib

\begin{thebibliography}{10}

\bibitem{adeSimonsObservatoryScience2019}
Ade, P., Aguirre, J., Ahmed, Z., Aiola, S., Ali, A., Alonso, D., Alvarez,
  M.~A., Arnold, K., Ashton, P., Austermann, J., Awan, H., Baccigalupi, C.,
  Baildon, T., Barron, D., Battaglia, N., Battye, R., Baxter, E., Bazarko, A.,
  Beall, J.~A., Bean, R., Beck, D., Beckman, S., Beringue, B., Bianchini, F.,
  Boada, S., Boettger, D., Bond, J.~R., Borrill, J., Brown, M.~L., Bruno,
  S.~M., Bryan, S., Calabrese, E., Calafut, V., Calisse, P., Carron, J.,
  Challinor, A., Chesmore, G., Chinone, Y., Chluba, J., Cho, H.-M.~S., Choi,
  S., Coppi, G., Cothard, N.~F., Coughlin, K., Crichton, D., Crowley, K.~D.,
  Crowley, K.~T., Cukierman, A., D'Ewart, J.~M., D{\"u}nner, R., de~Haan, T.,
  Devlin, M., Dicker, S., Didier, J., Dobbs, M., Dober, B., Duell, C.~J., Duff,
  S., Duivenvoorden, A., Dunkley, J., Dusatko, J., Errard, J., Fabbian, G.,
  Feeney, S., Ferraro, S., Flux{\`a}, P., Freese, K., Frisch, J.~C., Frolov,
  A., Fuller, G., Fuzia, B., Galitzki, N., Gallardo, P.~A., Ghersi, J. T.~G.,
  Gao, J., Gawiser, E., Gerbino, M., Gluscevic, V., {Goeckner-Wald}, N., Golec,
  J., Gordon, S., Gralla, M., Green, D., Grigorian, A., Groh, J., Groppi, C.,
  Guan, Y., Gudmundsson, J.~E., Han, D., Hargrave, P., Hasegawa, M.,
  Hasselfield, M., Hattori, M., Haynes, V., Hazumi, M., He, Y., Healy, E.,
  Henderson, S.~W., {Hervias-Caimapo}, C., Hill, C.~A., Hill, J.~C., Hilton,
  G., Hilton, M., Hincks, A.~D., Hinshaw, G., Hlo{\v z}ek, R., Ho, S., Ho,
  S.-P.~P., Howe, L., Huang, Z., Hubmayr, J., Huffenberger, K., Hughes, J.~P.,
  Ijjas, A., Ikape, M., Irwin, K., Jaffe, A.~H., Jain, B., Jeong, O., Kaneko,
  D., Karpel, E.~D., Katayama, N., Keating, B., Kernasovskiy, S.~S., Keskitalo,
  R., Kisner, T., Kiuchi, K., Klein, J., Knowles, K., Koopman, B., Kosowsky,
  A., Krachmalnicoff, N., Kuenstner, S.~E., Kuo, C.-L., Kusaka, A., Lashner,
  J., Lee, A., Lee, E., Leon, D., Leung, J. S.-Y., Lewis, A., Li, Y., Li, Z.,
  Limon, M., Linder, E., {Lopez-Caraballo}, C., Louis, T., Lowry, L., Lungu,
  M., Madhavacheril, M., Mak, D., Maldonado, F., Mani, H., Mates, B., Matsuda,
  F., Maurin, L., Mauskopf, P., May, A., McCallum, N., McKenney, C., McMahon,
  J., Meerburg, P.~D., Meyers, J., Miller, A., Mirmelstein, M., Moodley, K.,
  Munchmeyer, M., Munson, C., Naess, S., Nati, F., Navaroli, M., Newburgh, L.,
  Nguyen, H.~N., Niemack, M., Nishino, H., {Orlowski-Scherer}, J., Page, L.,
  Partridge, B., Peloton, J., Perrotta, F., Piccirillo, L., Pisano, G.,
  Poletti, D., Puddu, R., Puglisi, G., Raum, C., Reichardt, C.~L., Remazeilles,
  M., Rephaeli, Y., Riechers, D., Rojas, F., Roy, A., Sadeh, S., Sakurai, Y.,
  Salatino, M., Rao, M.~S., Schaan, E., Schmittfull, M., Sehgal, N., Seibert,
  J., Seljak, U., Sherwin, B., Shimon, M., Sierra, C., Sievers, J., Sikhosana,
  P., {Silva-Feaver}, M., Simon, S.~M., Sinclair, A., Siritanasak, P., Smith,
  K., Smith, S.~R., Spergel, D., Staggs, S.~T., Stein, G., Stevens, J.~R.,
  Stompor, R., Suzuki, A., Tajima, O., Takakura, S., Teply, G., Thomas, D.~B.,
  Thorne, B., Thornton, R., Trac, H., Tsai, C., Tucker, C., Ullom, J.,
  Vagnozzi, S., van Engelen, A., Lanen, J.~V., Winkle, D. D.~V., Vavagiakis,
  E.~M., Verg{\`e}s, C., Vissers, M., Wagoner, K., Walker, S., Ward, J.,
  Westbrook, B., Whitehorn, N., Williams, J., Williams, J., Wollack, E.~J., Xu,
  Z., Yu, B., Yu, C., Zago, F., Zhang, H., and {and}, N.~Z., ``The {{Simons
  Observatory}}: Science goals and forecasts,'' {\em Journal of Cosmology and
  Astroparticle Physics}~{\bf 2019},  056--056 (Feb. 2019).

\bibitem{zhuSimonsObservatoryLarge2021}
Zhu, N., Bhandarkar, T., Coppi, G., Kofman, A.~M., {Orlowski-Scherer}, J.~L.,
  Xu, Z., Adachi, S., Ade, P., Aiola, S., Austermann, J., Bazarko, A.~O.,
  Beall, J.~A., Bhimani, S., Bond, J.~R., Chesmore, G.~E., Choi, S.~K.,
  Connors, J., Cothard, N.~F., Devlin, M., Dicker, S., Dober, B., Duell, C.~J.,
  Duff, S.~M., D{\"u}nner, R., Fabbian, G., Galitzki, N., Gallardo, P.~A.,
  Golec, J.~E., Haridas, S.~K., Harrington, K., Healy, E., Ho, S.-P.~P., Huber,
  Z.~B., Hubmayr, J., Iuliano, J., Johnson, B.~R., Keating, B., Kiuchi, K.,
  Koopman, B.~J., Lashner, J., Lee, A.~T., Li, Y., Limon, M., Link, M., Lucas,
  T.~J., McCarrick, H., Moore, J., Nati, F., Newburgh, L.~B., Niemack, M.~D.,
  Pierpaoli, E., Randall, M.~J., Sarmiento, K.~P., Saunders, L.~J., Seibert,
  J., Sierra, C., Sonka, R., Spisak, J., Sutariya, S., Tajima, O., Teply,
  G.~P., Thornton, R.~J., Tsan, T., Tucker, C., Ullom, J., Vavagiakis, E.~M.,
  Vissers, M.~R., Walker, S., Westbrook, B., Wollack, E.~J., and Zannoni, M.,
  ``The {{Simons Observatory Large Aperture Telescope Receiver}},'' {\em The
  Astrophysical Journal Supplement Series}~{\bf 256},  23 (Sept. 2021).

\bibitem{parshleyOpticalDesignSixmeter2018}
Parshley, S.~C., Niemack, M., Hills, R., Dicker, S.~R., D{\"u}nner, R., Erler,
  J., Gallardo, P.~A., Gudmundsson, J.~E., Herter, T., Koopman, B.~J., Limon,
  M., Matsuda, F.~T., Mauskopf, P., Riechers, D.~A., Stacey, G.~J., and
  Vavagiakis, E.~M., ``The optical design of the six-meter {{CCAT-prime}} and
  {{Simons Observatory}} telescopes,'' in [{\em Ground-Based and {{Airborne
  Telescopes VII}}}{\nolinebreak\hspace{0.1em}]},   {\bf 10700},  1292--1304,
  {SPIE} (July 2018).

\bibitem{galitzkiSimonsObservatoryInstrument2018}
Galitzki, N., Ali, A., Arnold, K.~S., Ashton, P.~C., Austermann, J.~E.,
  Baccigalupi, C., Baildon, T., Barron, D., Beall, J.~A., Beckman, S., Bruno,
  S. M.~M., Bryan, S., Calisse, P.~G., Chesmore, G.~E., Chinone, Y., Choi,
  S.~K., Coppi, G., Crowley, K.~D., Crowley, K.~T., Cukierman, A., Devlin,
  M.~J., Dicker, S., Dober, B., Duff, S.~M., Dunkley, J., Fabbian, G.,
  Gallardo, P.~A., Gerbino, M., {Goeckner-Wald}, N., Golec, J.~E., Gudmundsson,
  J.~E., Healy, E.~E., Henderson, S., Hill, C.~A., Hilton, G.~C., Ho, S.-P.~P.,
  Howe, L.~A., Hubmayr, J., Jeong, O., Keating, B., Koopman, B.~J., Kuichi, K.,
  Kusaka, A., Lashner, J., Lee, A.~T., Li, Y., Limon, M., Lungu, M., Matsuda,
  F., Mauskopf, P.~D., May, A.~J., McCallum, N., McMahon, J., Nati, F.,
  Niemack, M.~D., {Orlowski-Scherer}, J.~L., Parshley, S.~C., Piccirillo, L.,
  Rao, M.~S., Raum, C., Salatino, M., Seibert, J.~S., Sierra, C.,
  {Silva-Feaver}, M., Simon, S.~M., Staggs, S.~T., Stevens, J.~R., Suzuki, A.,
  Teply, G., Thornton, R., Tsai, C., Ullom, J.~N., Vavagiakis, E.~M., Vissers,
  M.~R., Westbrook, B., Wollack, E.~J., Xu, Z., and Zhu, N., ``The {{Simons
  Observatory}}: {{Instrument Overview}},'' {\em Millimeter, Submillimeter, and
  Far-Infrared Detectors and Instrumentation for Astronomy IX} ,  3 (July
  2018).

\bibitem{aliSmallApertureTelescopes2020}
Ali, A.~M., Adachi, S., Arnold, K., Ashton, P., Bazarko, A., Chinone, Y.,
  Coppi, G., Corbett, L., Crowley, K.~D., Crowley, K.~T., Devlin, M., Dicker,
  S., Duff, S., Ellis, C., Galitzki, N., {Goeckner-Wald}, N., Harrington, K.,
  Healy, E., Hill, C.~A., Ho, S.-P.~P., Hubmayr, J., Keating, B., Kiuchi, K.,
  Kusaka, A., Lee, A.~T., Ludlam, M., Mangu, A., Matsuda, F., McCarrick, H.,
  Nati, F., Niemack, M.~D., Nishino, H., {Orlowski-Scherer}, J.,
  Sathyanarayana~Rao, M., Raum, C., Sakurai, Y., Salatino, M., Sasse, T.,
  Seibert, J., Sierra, C., {Silva-Feaver}, M., Spisak, J., Simon, S.~M.,
  Staggs, S., Tajima, O., Teply, G., Tsan, T., Wollack, E., Westbrook, B., Xu,
  Z., Zannoni, M., and Zhu, N., ``Small {{Aperture Telescopes}} for the
  {{Simons Observatory}},'' {\em Journal of Low Temperature Physics}~{\bf 200},
   461--471 (Sept. 2020).

\bibitem{matesDemonstrationMultiplexerDissipationless2008}
Mates, J. a.~B., Hilton, G.~C., Irwin, K.~D., Vale, L.~R., and Lehnert, K.~W.,
  ``Demonstration of a multiplexer of dissipationless superconducting quantum
  interference devices,'' {\em Applied Physics Letters}~{\bf 92},  023514 (Jan.
  2008).

\bibitem{hendersonHighlymultiplexedMicrowaveSQUID2018}
Henderson, S.~W., Ahmed, Z., Austermann, J., Becker, D., Bennett, D.~A., Brown,
  D., Chaudhuri, S., Cho, H.-M.~S., D'Ewart, J.~M., Dober, B., Duff, S.~M.,
  Dusatko, J.~E., Fatigoni, S., Frisch, J.~C., Gard, J.~D., Halpern, M.,
  Hilton, G.~C., Hubmayr, J., Irwin, K.~D., Karpel, E.~D., Kernasovskiy, S.~S.,
  Kuenstner, S.~E., Kuo, C.-L., Li, D., Mates, J. A.~B., Reintsema, C.~D.,
  Smith, S.~R., Ullom, J., Vale, L.~R., Van~Winkle, D.~D., Vissers, M., and Yu,
  C., ``Highly-multiplexed microwave {{SQUID}} readout using the {{SLAC
  Microresonator Radio Frequency}} ({{SMuRF}}) {{Electronics}} for {{Future
  CMB}} and {{Sub-millimeter Surveys}},'' {\em Millimeter, Submillimeter, and
  Far-Infrared Detectors and Instrumentation for Astronomy IX} ,  43 (July
  2018).

\bibitem{mccarrickSimonsObservatoryMicrowave2021}
McCarrick, H., Healy, E., Ahmed, Z., Arnold, K., Atkins, Z., Austermann, J.~E.,
  Bhandarkar, T., Beall, J.~A., Bruno, S.~M., Choi, S.~K., Connors, J.,
  Cothard, N.~F., Crowley, K.~D., Dicker, S., Dober, B., Duell, C.~J., Duff,
  S.~M., Dutcher, D., Frisch, J.~C., Galitzki, N., Gralla, M.~B., Gudmundsson,
  J.~E., Henderson, S.~W., Hilton, G.~C., Ho, S.-P.~P., Huber, Z.~B., Hubmayr,
  J., Iuliano, J., Johnson, B.~R., Kofman, A.~M., Kusaka, A., Lashner, J., Lee,
  A.~T., Li, Y., Link, M.~J., Lucas, T.~J., Lungu, M., Mates, J. A.~B.,
  McMahon, J.~J., Niemack, M.~D., {Orlowski-Scherer}, J., Seibert, J.,
  {Silva-Feaver}, M., Simon, S.~M., Staggs, S., Suzuki, A., Terasaki, T.,
  Thornton, R., Ullom, J.~N., Vavagiakis, E.~M., Vale, L.~R., Lanen, J.~V.,
  Vissers, M.~R., Wang, Y., Wollack, E.~J., Xu, Z., Young, E., Yu, C., Zheng,
  K., and Zhu, N., ``The {{Simons Observatory Microwave SQUID Multiplexing
  Detector Module Design}},'' {\em The Astrophysical Journal}~{\bf 922},  38
  (Nov. 2021).

\bibitem{matesFluxRampModulationSQUID2012}
Mates, J. A.~B., Irwin, K.~D., Vale, L.~R., Hilton, G.~C., Gao, J., and
  Lehnert, K.~W., ``Flux-{{Ramp Modulation}} for {{SQUID Multiplexing}},'' {\em
  Journal of Low Temperature Physics}~{\bf 167},  707--712 (June 2012).

\bibitem{wangSimonsObservatoryFocalPlane2021}
Wang, Y., Zheng, K., Atkins, Z., Austermann, J., Bhandarkar, T., Choi, S.~K.,
  Duff, S.~M., Dutcher, D., Galitzki, N., Healy, E., Huber, Z.~B., Hubmayr, J.,
  Johnson, B.~R., Lashner, J., Li, Y., McCarrick, H., Niemack, M.~D., Seibert,
  J., Staggs, S.~T., Vavagiakis, E., and Xu, Z., ``Simons {{Observatory
  Focal-Plane Module}}: {{In-lab Testing}} and {{Characterization Program}},''
  {\em arXiv:2111.11301 [astro-ph, physics:physics]}  (Nov. 2021).

\bibitem{cothardComparingComplexImpedance2020}
Cothard, N.~F., Ali, A.~M., Austermann, J.~E., Choi, S.~K., Crowley, K.~T.,
  Dober, B.~J., Duell, C.~J., Duff, S.~M., Gallardo, P., Hilton, G.~C., Ho,
  S.-P.~P., Hubmayr, J., Link, M.~J., Niemack, M.~D., Sonka, R.~F., Staggs,
  S.~T., Vavagiakis, E.~M., Wollack, E.~J., and Xu, Z., ``Comparing complex
  impedance and bias step measurements of {{Simons Observatory}} transition
  edge sensors,'' in [{\em Millimeter, {{Submillimeter}}, and {{Far-Infrared
  Detectors}} and {{Instrumentation}} for {{Astronomy
  X}}}{\nolinebreak\hspace{0.1em}]},   {\bf 11453},  302--311, {SPIE} (Dec.
  2020).

\bibitem{irwinTransitionEdgeSensors2005}
Irwin, K. and Hilton, G., ``Transition-{{Edge Sensors}},'' in [{\em Cryogenic
  {{Particle Detection}}}{\nolinebreak\hspace{0.1em}]},  Enss, C., ed., {\em
  Topics in {{Applied Physics}}},  63--150, {Springer}, {Berlin, Heidelberg}
  (2005).

\bibitem{beckerSUBMILLIMETERVIDEOIMAGING}
Becker, D.~T., {\em {{SUBMILLIMETER VIDEO IMAGING WITH A SUPERCONDUCTING
  BOLOMETER ARRAY}}}, PhD thesis.

\bibitem{seibertDevelopmentOpticalDetector2020}
Seibert, J., Ade, P., Ali, A.~M., Arnold, K., Cothard, N.~F., Galitzki, N.,
  Harrington, K., Ho, S.-P.~P., Keating, B., Lowry, L.~N., Russell, M.,
  {Silva-Feaver}, M., Siritanasak, P., Teply, G.~P., Tucker, C., Vavagiakis,
  E.~M., and Xu, Z., ``Development of an optical detector testbed for the
  {{Simons Observatory}},'' in [{\em Millimeter, {{Submillimeter}}, and
  {{Far-Infrared Detectors}} and {{Instrumentation}} for {{Astronomy
  X}}}{\nolinebreak\hspace{0.1em}]},   {\bf 11453},  114532C, {International
  Society for Optics and Photonics} (Dec. 2020).

\bibitem{probstLowFrequencyDigital1985}
Probst, P.~A. and Collet, B., ``Low-frequency digital lock-in amplifier,'' {\em
  Review of Scientific Instruments}~{\bf 56},  466--470 (Mar. 1985).

\bibitem{lindemanComplexImpedanceMeasurements2007}
Lindeman, M.~A., Barger, K.~A., Brandl, D.~E., Crowder, S.~G., Rocks, L., and
  McCammon, D., ``Complex impedance measurements of calorimeters and
  bolometers: {{Correction}} for stray impedances,'' {\em Review of Scientific
  Instruments}~{\bf 78},  043105 (Apr. 2007).

\bibitem{koopmanAdvancedACTPolLowFrequency2018}
Koopman, B.~J., Cothard, N.~F., Choi, S.~K., Crowley, K.~T., Duff, S.~M.,
  Henderson, S.~W., Ho, S.~P., Hubmayr, J., Gallardo, P.~A., Nati, F., Niemack,
  M.~D., Simon, S.~M., Staggs, S.~T., Stevens, J.~R., Vavagiakis, E.~M., and
  Wollack, E.~J., ``Advanced {{ACTPol Low-Frequency Array}}: {{Readout}} and
  {{Characterization}} of {{Prototype}} 27 and 39~{{GHz Transition Edge
  Sensors}},'' {\em Journal of Low Temperature Physics}~{\bf 193},  1103--1111
  (Dec. 2018).

\bibitem{niemackDarkEnergyDesign2008}
Niemack, M.~D., {\em Towards Dark Energy: {{Design}}, Development, and
  Preliminary Data from {{ACT}}}, PhD thesis (Jan. 2008).

\end{thebibliography}
\bibliographystyle{spiebib} % makes bibtex use spiebib.bst

\end{document}